\documentclass[]{spie}  

\newcommand{\z}{\mathbf{z}}
\newcommand{\f}{\mathbf{f}}
\newcommand{\w}{\mathbf{w}}

\newcommand{\g}{\mathbf{g}}
\newcommand{\R}{\mathbb{R}}

\newcommand\numberthis{\addtocounter{equation}{1}\tag{\theequation}}

\renewcommand{\u}{\mathbf{u}}

\providecommand{\norm}[1]{\ensuremath{\left\lVert#1\right\rVert}}

\def\[{\begin{equation}}
\def\]{\end{equation}}

\newcommand{\fpi}{\ensuremath{\f^{(\rm{PI})}}}
\newcommand{\wpi}{\ensuremath{\w^{(\rm{PI})}}}
 
\usepackage{amsmath,amsfonts,amssymb}
\usepackage{graphicx}
\usepackage{subcaption}
\usepackage{placeins}
\usepackage{cite}
\usepackage{booktabs}
\usepackage[colorlinks=true, allcolors=blue]{hyperref}

\title{Prior image-based medical image reconstruction using a style-based generative adversarial network}

\author[a]{Varun A. Kelkar}
\author[a,b]{Mark A. Anastasio}

\affil[a]{Department of Elect. \& Computer Eng., University of Illinois at Urbana-Champaign, IL, USA}
\affil[b]{Department of Bioengineering, University of Illinois at Urbana-Champaign, IL 61801, USA}
\authorinfo{Send correspondence to Mark A. Anastasio. E-mail: maa@illinois.edu, Telephone: +1 (217) 300 0314}

\begin{document} 
\maketitle

\begin{abstract}
Computed medical imaging systems require a computational reconstruction procedure for image formation. In order to recover a useful estimate of the object to-be-imaged when the recorded measurements are incomplete, prior knowledge about the nature of object must be utilized. 
In order to improve the conditioning of an ill-posed imaging inverse problem, deep learning approaches are being actively investigated for better representing object priors and constraints.
This work proposes to use a style-based generative adversarial network (StyleGAN) to constrain an image reconstruction problem in the case where additional information in the form of a prior image of the sought-after object is available. An optimization problem is formulated in the intermediate latent-space of a StyleGAN, that is disentangled with respect to meaningful image attributes or ``styles", such as the contrast used in magnetic resonance imaging (MRI). Discrepancy between the sought-after and prior images is measured in the disentangled latent-space, and is used to regularize the inverse problem in the form of constraints on specific styles of the disentangled latent-space. A stylized numerical study inspired by MR imaging is designed, where the sought-after and the prior image are structurally similar, but belong to different contrast mechanisms. The presented numerical studies demonstrate the superiority of the proposed approach as compared to classical approaches in the form of traditional metrics. 
\end{abstract}

\keywords{Inverse problems; generative adversarial network; compressive sensing; reference-based MRI}

\section{Purpose}
\label{sec:purpose}
Many medical imaging systems are well approximated by a discrete-to-discrete linear imaging model described as \cite{barrett}
\begin{align}\label{eqn:img_system}
    \g = H\f + \mathbf{n},
\end{align}
where $\f \in \mathbb{E}^n$ corresponds to the object to-be-imaged, $\g \in \mathbb{E}^m$ corresponds to the measurements taken, $H \in \mathbb{E}^{m\times n}$ corresponds to the linear operator that approximately describes the imaging system, and $\mathbf{n} \in \mathbb{E}^m$ is the measurement noise. Estimating $\f$ from $\g$ represents a linear inverse problem, which is often ill-posed. Hence, prior knowledge about the nature of $\f$ needs to be imposed in order to obtain a useful estimate of $\f$. A traditional class of techniques for obtaining an estimate of $\f$ involves formulating the inverse problem in terms of an optimization problem:
\begin{align}
    \hat{\f} = \arg\min_\f \|\g - H\f\|_2^2 + \lambda \Phi(\f),
\end{align}
where $\Phi(\f)$ denotes the regularization that imposes prior knowledge about $\f$ in the form of a penalty, and $\lambda$ is the regularization parameter. Sparsity-promoting penalties have been used to successfully estimate $\f$ when $\f$ is sparse in some domain, and $H$ is incoherent with the sparsifying transform \cite{candes_review, sparsemri}.

However, sparsity-promoting priors may not be able to fully capture the complex distribution of medical images. In recent years, deep generative models, such as generative adversarial networks (GANs), have shown great promise in learning distributions of objects \cite{goodfellow, progan}, which can be employed as an object prior in an image reconstruction approach. These may potentially outperform traditional sparsity-based priors. For example, Bora \textit{et al.} proposed an approach in which the solution of a least squares image reconstruction problem is constrained to reside within the range of a generative model \cite{bora}, while promoting solutions that are consistent with the measurements. Over the years, there has been a tremendous improvement in the performance of GANs \cite{progan, stylegan, stylegan2}. Modern GANs such as StyleGAN and StyleGAN2 are able to generate realistic images, with a great degree of controllability over individual semantic image features, while also being able to better represent images outside the training data. 

In this work, style-based generative adversarial networks
are investigated for their ability to regularize a \textit{prior image-constrained reconstruction} problem where the desired object estimate is known to be related to a given prior image of the object.
This scenario is relevant when the same object is to be imaged twice with different contrast mechanisms, in longitudinal studies \cite{piccs}, or multi-contrast magnetic resonance imaging (MRI) \cite{refmri}. 
A traditional approach to this problem is to assume that the difference between the sought-after and the prior image is sparse in a linear transform domain \cite{piccs, cspi}, and to regularize the inverse problem by penalizing this difference. However, in many scenarios such as multi-contrast MR imaging, this assumption is not true. Hence, such a strategy may not be appropriate for ensuring that the sought-after image is related to the prior image in a meaningful way. Style-based GANs have the ability to individually control meaningful image attributes, or ``styles", in an image by representing the image in a disentangled latent space. {In this work}, prior image-constrained reconstruction is formulated as an optimization problem using the disentangled latent space of a style-based GAN. Prior image-based regularization is imposed by constraining the estimate to have certain styles equal to the corresponding styles of the prior image.

\section{Methods}
\subsection{Style-based GANs}
StyleGAN \cite{stylegan} and StyleGAN2 \cite{stylegan2} are characterized by an generator architecture consisting of two networks - (1) a \textit{mapping network} $g_{\rm{mapping}} : \R^{k} \rightarrow \R^{k}$, and an $L$-layer \textit{synthesis network} $G : \R^{Lk} \rightarrow \R^n$. During conventional image generation, $g_{\rm mapping}$ maps a sample $\z \in \mathcal{Z} \equiv \R^k$ from an iid standard normal distribution to $\u \in \mathcal{W} \equiv \R^k$. Subsequently, $L$ copies of $\u$ are stacked in order to form a $K = kL$ dimensional vector $\w \in \mathcal{W}^+ \equiv \R^K$, that is utilized as input to the $L$-layer synthesis network $G$. The $i$th level of detail in the image is controlled by the $i$th copy of $\u$. This architecture, along with a specific regularization scheme introduced in the StyleGAN2 paper \cite{stylegan}, lends the StyleGAN the ability to control individual semantic features at different scales. StyleGAN2 also takes advantage of the path-length regularization, which aids in better conditioning of $G$ and reducing the representation error \cite{stylegan2}. In the conventional image generation scheme, an image sampled from StyleGAN2 corresponds to a degenerate $\w$ vector, containing $L$ copies of $\u$. However, from the point-of-view of finding an accurate latent representation of a given image, utilizing the entire $\mathcal{W}^+$ space leads to lower representation error \cite{stylegan_inv_gaussian}.

\begin{figure}[t]
\centering
\includegraphics[width=\textwidth]{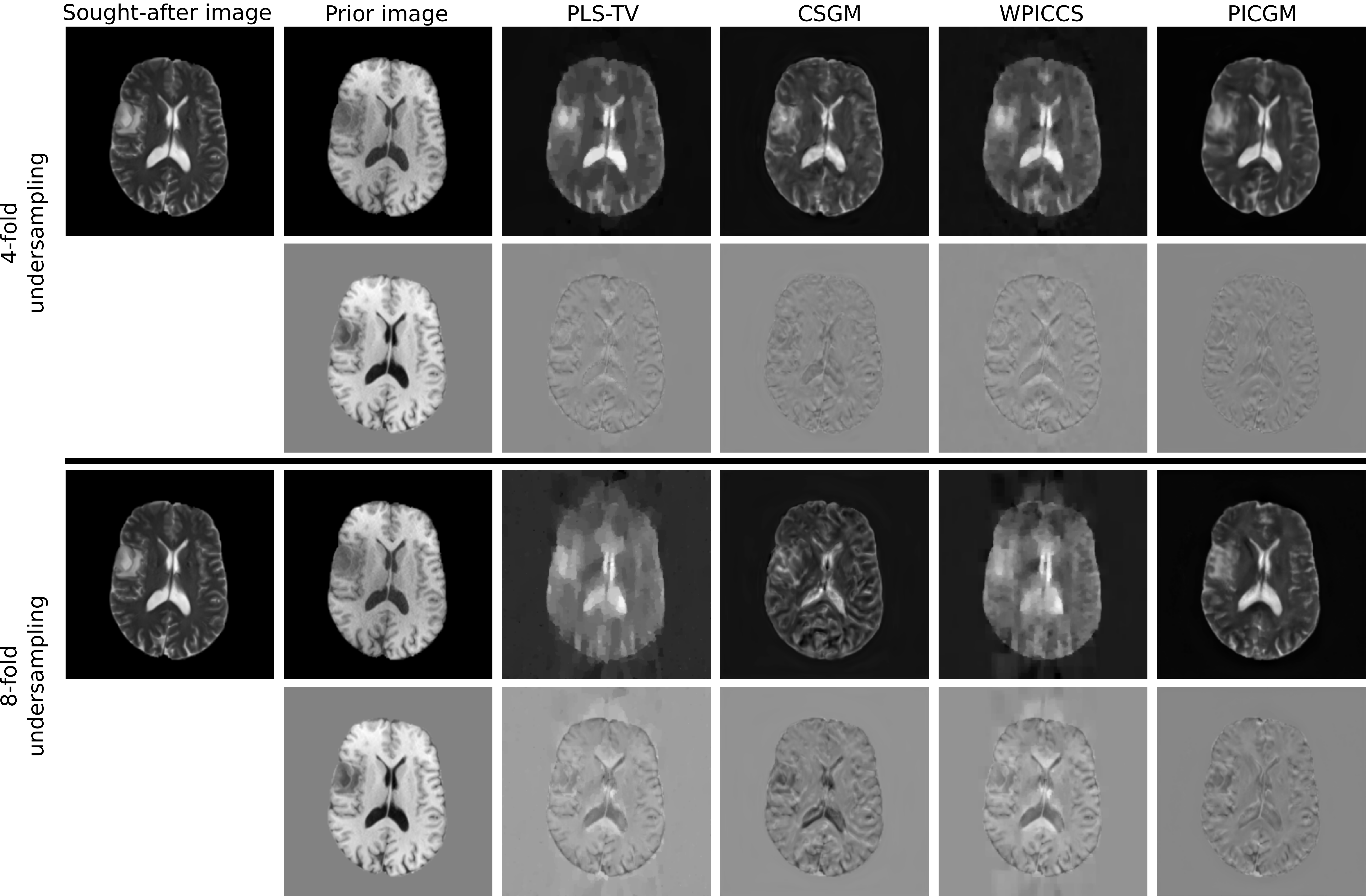}
\vspace{1pt}
\captionof{figure}{Sought-after image, prior image and image reconstruction results for an axial brain image, for 4-fold and 8-fold Cartesian undersampling and 20 dB SNR. For each undersampling case, the lower row of images shows the difference image with respect to the sought-after image.}
\vspace{-5pt}
\label{fig:brats_recon}
\end{figure}

\subsection{Prior image-constrained reconstruction using StyleGAN2}
StyleGAN and StyleGAN2 can keep certain styles of an image fixed, while varying certain other styles. For a medical image dataset, such as a dataset of multi-contrast brain MRI images, a StyleGAN can control the contrast, or the placement of the ventricles, folds, or other fine scale features, while keeping the general structure of the image the same. Therefore, a natural way of quantifying the similarity between the sought-after and the prior images is to compare their disentangled latent-representations.

\begin{figure}[t]
\centering
\includegraphics[width=0.65\textwidth]{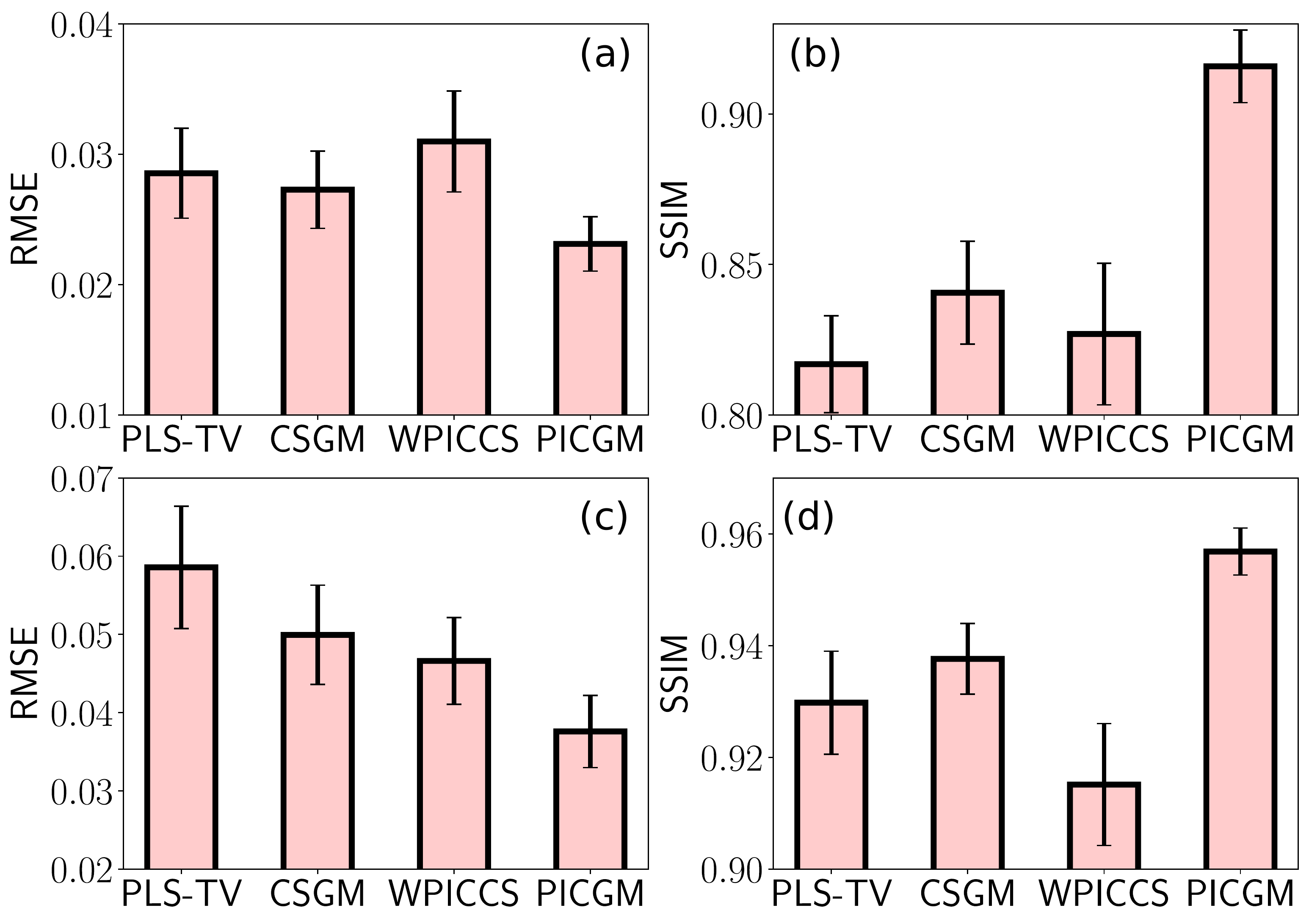}
\vspace{1pt}
\captionof{figure}{RMSE and SSIM values, respectively, for (a,b) four-fold and (c,d) 8-fold Cartesian undersampling}
\vspace{5pt}
\label{fig:rmses_ssims}
\end{figure}

\begin{figure}[t]
\centering
\includegraphics[width=0.8\textwidth]{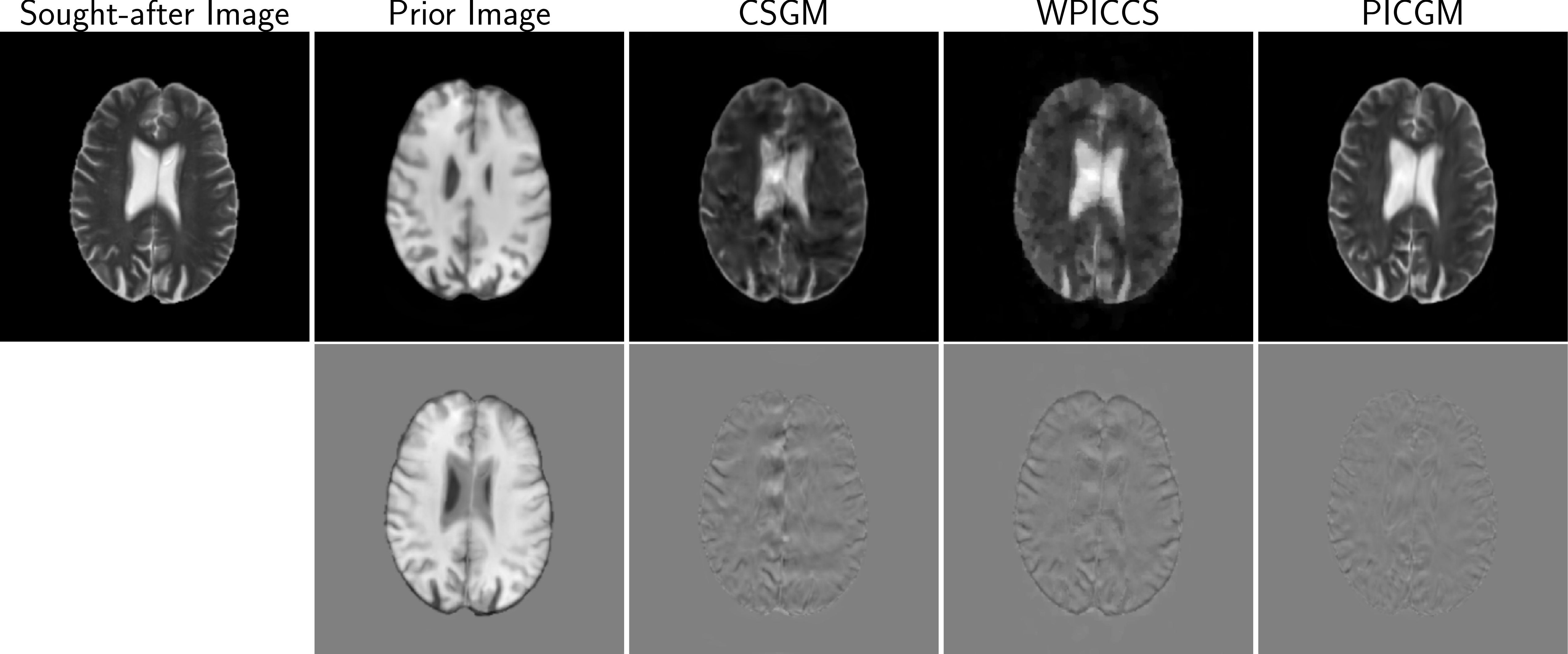}
\vspace{1pt}
\captionof{figure}{Sought-after image, prior image and image reconstruction results for an axial brain image, for 4-fold Cartesian undersampling and 20 dB SNR for the case where the sought-after and prior images are misaligned.}
\vspace{-5pt}
\label{fig:brats_recon_robust}
\end{figure}

\begin{figure}[t]
\centering
\includegraphics[width=0.65\textwidth]{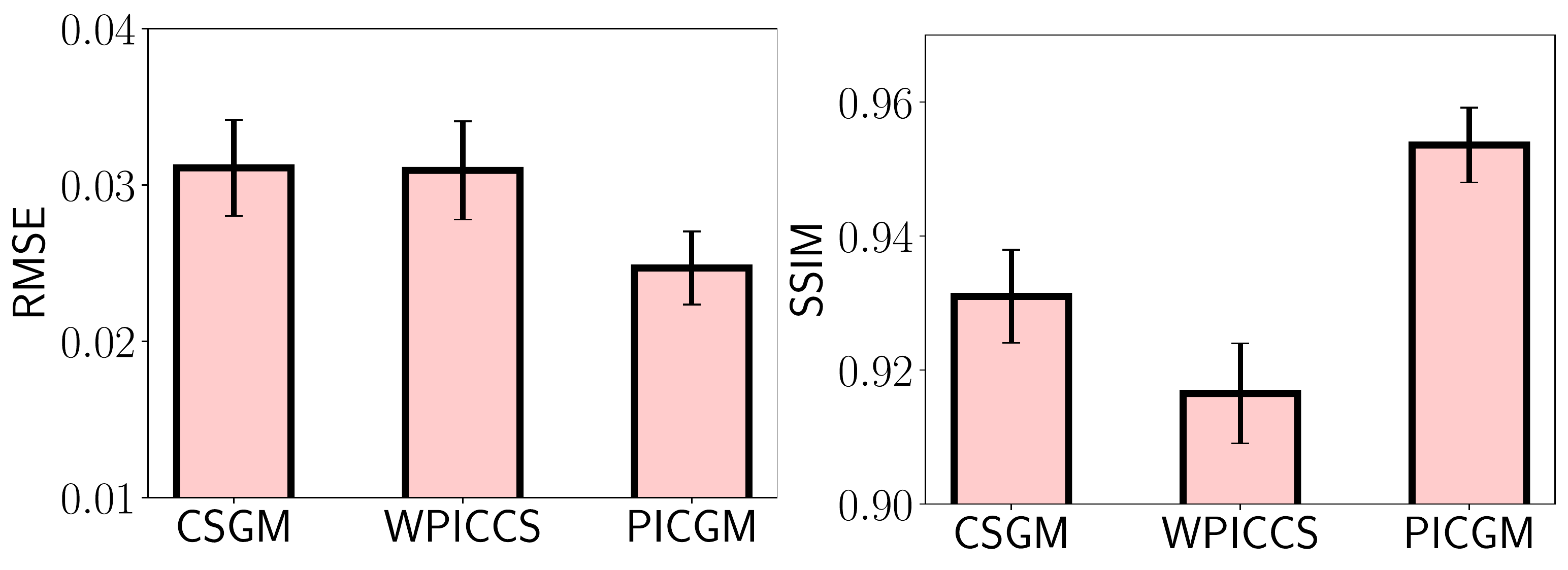}
\vspace{1pt}
\captionof{figure}{RMSE and SSIM values for 4-fold Cartesian undersampling when the sought-after and prior images are misaligned.}
\vspace{-5pt}
\label{fig:robust_rmses_ssims}
\end{figure}

Taking inspiration from the style-mixing properties of a StyleGAN, the prior image-constrained inverse problem can be formulated as follows. Let $G : \mathbb{R}^K \rightarrow \mathbb{R}^n$ denote the synthesis network of a trained StyleGAN2. Let $\fpi = G(\wpi)$ be the known prior image in the range of $G$. Let $\tilde{\f} \in \{G(\w) ~ s.t. ~ \w_{1:p_1} = \wpi_{1:p_1}, \w_{p_2:K} = \wpi_{p_2:K}\}$ represent the sought-after image. Here, $\w_{u:v}$ denotes the section of vector $\w$ from indices $u$ through $v$. Then, the proposed reconstruction framework can be expressed in terms of the following optimization problem:
\begin{align*}
    \hat{\w} &= \arg\min_{\w} \norm{\g - HG(\w)}_2^2 + \lambda\phi(\w),\\
    \text{s.t. } \w_{1:p_1} &= \wpi_{1:p_1}, \quad \w_{p_2:K} = \wpi_{p_2:K};\\
    \hat{\f} &= G(\hat{\w}).\numberthis{}\label{eqn:picgm}
\end{align*}
where $p_1, p_2$ are multiples of $k$, $1 \leq p_1 < p_2 \leq K$. Here, $\phi(\w)$ is a regularization term formulated based on the Gaussianization of latent space, discussed in the recent literature \cite{stylegan_inv_gaussian}.
Although the above problem is non-convex, useful estimates $\hat{\f}$ can be obtained by gradient descent-based optimization similar to previous works \cite{bora, aliahmed}. A projected-Adam algorithm is used for this purpose \cite{adam}. Note that a similar strategy of constraining sections of the latent space of a generative model to regularize generative model-constrained image reconstruction has been explored previously \cite{clinn}.

\subsection{Weighted prior image-constrained compressed sensing (WPICCS)} \label{sec:wpiccs}
The proposed method is compared against a classical approach that assumes sparsity in a weighted different between the sought-after and prior images \cite{refmri}. This approach is referred to here as weighted prior image-constrained compressed sensing (WPICCS). In this setting, the reconstruction framework can be described in terms of the following optimization problem:
\begin{align}\label{eqn:wpiccs}
    \hat{\f} = \arg\min_{\f} \norm{ \g - H\f }_2^2 + \lambda\left(\alpha\norm{W\Psi(\f - \fpi)}_1 + (1-\alpha)\norm{\Phi\f}_1 \right).
\end{align}
Here, $\Phi$ is the sparsifying transform under which $\tilde{\f}$ is expected to be compressible, $\Psi$ is the sparsifying transform under which the difference $\tilde{\f} - \fpi$ is expected to be compressible, and $W$ is a diagonal matrix that weights the transformed components. In practice, Eq. (\ref{eqn:wpiccs}) can be solved using an iterative minimization scheme, that also updates $W$ periodically \cite{refmri}. 


\section{Numerical studies}
In this study, the proposed method is compared with the traditional penalized least squares method with TV regularization (PLS-TV), compressed sensing with generative models (CSGM) \cite{bora} using StyleGAN2 and the WPICCS method described in Section \ref{sec:wpiccs}.

For CSGM and PICGM, the StyleGAN2 was trained on 70000 axial slices of size 256$\times$256 from T1 and T2 weighted brain MRI scans from the BraTS dataset \cite{brats}. The StyleGAN2 was trained using 4 11 GB NVIDIA TITAN X (Pascal) GPUs for a period of around 4 days using Tensorflow 1.14/Python 3.7. The default settings for the training parameters were employed.

For evaluating reconstruction performance, the measurements were \textit{stylized} single-coil MR measurements produced from a T2-weighted axial brain image. For WPICCS and PICGM, the prior image was a T1 weighted image, whereas corresponding to the matching slice of the same volume. Cartesian random undersampling masks corresponding to R = 4 and R = 8 undersampling ratios were utilized \cite{fastmri}. All the regularization parameters were tuned with the help of grid search on a single validation image, and the regularization parameter configurations giving the lowest mean squared error (MSE) were chosen. They were then used to perform image reconstruction for a test dataset of size 20. For WPICCS, the wavelet transform with 7 levels was chosen as the sparsifying transform $\Psi$ and the discrete difference operator (corresponding to the TV-seminorm) was used as the sparsifying transform $\Phi$.

Lastly, the ability of the proposed approach to recover images in the case where the sought-after and prior images are not perfectly matched, was studied. For each of the sought-after T2-weighted axial brain images from the test dataset, a T1-weighted image located 5 axial slices from the paired T1-weighted image, towards or away from the skull roof with equal probability, was chosen as the prior image. Image reconstruction from four-fold Cartesian undersampled measurements with 20 dB SNR was performed by use of the WPICCS, CSGM and PICGM methods with configurations described above, and image quality was compared.

\section{Results}
Figure \ref{fig:brats_recon} shows the images reconstructed using the various methods, from 4 fold and 8 fold simulated undersampled MRI measurements. The root-mean-square error (RMSE) and structural similarity (SSIM) values over a test dataset are shown in Fig. \ref{fig:rmses_ssims}. For the case of 8-fold undersampling, although the images recovered by PLS-TV, CSGM and WPICCS all have artefacts, CSGM has noticeably better defined boundaries. The proposed method outperforms all the other approaches examined in terms of RMSE and SSIM. Notably, the hyperparameters $p_1$ and $p_2$ which control the styles that are optimized over remain robust across the test dataset.

Additionally, it can be seen that the proposed approach is robust to misalignments in the sought-after and prior images such as unmatched slice locations. As shown in Fig. \ref{fig:brats_recon_robust}, PICGM is also able to recover the ventricles well, which is the region where the sought-after and prior images qualitatively differ the most. Quantitatively, Fig. \ref{fig:robust_rmses_ssims} shows that the PICGM outperforms the other approaches examined even in the case of misaligned sought-after and prior images.


\section{Conclusions}
In conclusion, training of a novel StyleGAN2 on MR images from the BraTS dataset was achieved and a novel formulation of the classical prior-image constrained reconstruction was developed. It was shown that the proposed approach gives superior qualitative and quantitative performance in terms of traditional image quality metrics as compared to classical approaches, even in the case where the sought-after and prior images are not perfectly matched. It should be noted that the generalization capabilities of generative model-constrained reconstruction requires further investigation, due to the possibility of imaging hallucinations that might occur due to mismatch of the prior \cite{hallucinations}. Lastly, we emphasize that a task-based assessment of image reconstruction methods is needed in order to translate these methods into the clinic \cite{barrett, jha, evalsr}.

\section{Disclosure}
Aspects of this work were published as a part of the International Conference on Machine Learning (ICML) 2021 \cite{picgm_icml}.

\acknowledgments
The authors would like to thank Sayantan Bhadra and Weimin Zhou for their help. This work was supported in part by NIH Awards EB020604, EB023045, NS102213, EB028652, and NSF Award DMS1614305.

\bibliography{report} 
\bibliographystyle{spiebib} 

\end{document}